# Design and Simulation of Capacitive Pressure Sensor for Blood Pressure Sensing Application

Rishabh Bhooshan Mishra, S. Santosh Kumar, Ravindra Mukhiya
Smart Sensors Area, CSIR - Central Electronics Engineering Research Institute, Pilani, Raj. India - 333031
santoshkumar.ceeri@gmail.com

**Abstract.** This paper presents the mathematical modeling based design and simulation of normal mode MEMS capacitive pressure sensor for blood pressure sensing application. The typical range of blood pressure is 80 – 120 mm Hg. But this range varies in case of any stress, hypertension and some other health issues. Analytical simulation is implemented using MATLAB®. Basically, normal mode capacitive pressure sensors have a fixed backplate and a moveable diaphragm which deflects on application of pressure with the condition that it must not touch the backplate. Deflection depends on material as well as thickness, shape and size of diaphragm which can be of circular, elliptical, square or rectangular shape. In this paper, circular shape is chosen due to higher sensitivity compared to other diaphragm shapes. Deflection, base capacitance, change in capacitance after applying pressure and sensitivity is reported for systolic and diastolic blood pressure monitoring application and study involves determining the optimized design for the sensor. Diaphragm deflection shows linear variation with applied pressure, which follows Hook's Law. The variation in capacitance is logarithmic function of applied pressure, which is utilized for analytical simulation.

## 1 Introduction

Micro Electro Mechanical System (MEMS) is an interesting and popular research area which widely combines the physical, chemical, biological processes on a chip [1]. The single integrated system or device can be analyzed by a specific partial differential equation based on domain of processes [2]. In all the MEMS devices, except microfluidics area, simple micromechanical and moveable structures like diaphragms and beams are used. These simple microstructures help in measuring the various process variables like pressure, humidity, acceleration, flow, temperature, pH etc. Micromachining and manufacturing science is the combination of tools and techniques which are adapted from VLSI technology and microelectronic fabrication [3]. In designing the pressure sensors for absolute, gauge or differential pressure measurement such as piezoresistive, capacitive etc., usually diaphragms are used. In piezoresistive pressure sensors, four piezoresistors are implanted on the diaphragm and when pres-sure is applied, diaphragm deflects. Due to change in resistance of piezoresistors, the Wheatstone bridge becomes unbalanced and an output voltage is obtained across the bridge. The applied pressure can be obtained according to the output voltage. The capacitive sensing of pressure offers various advantages over piezoresistive sensing such as higher sensitivity, low power consumption, long term stability and temperature insensitivity [4, 5].

Human body is combination of several instrumentation systems like cardiovascular system, respiratory system, nervous system, vascular system etc. A lot of research in Bio-MEMS for man-instrumentation system is in trend like blood pressure measurement, healthcare application, medical-diagnostics, liver on a chip and brain on a chip. Research in cardiovascular psychology and hypertension for various small mammals and human has been carried out using capacitive pressure sensors with new micromachining processes. In these sensors, the capacitive pressure sensors have been designed in an array which gives improved output signals and increases stability. A switched capacitor CMOS circuitry is used to measure the change in capacitance in the sensor array [6]. Therefore, using capacitive pressure sensors, the blood pressure measurement can be carried out. The normal mode capacitive pressure sensor has two parallel plates which are separated by a medium, one of the plates is diaphragm and the other is a fixed plate. After application of pressure, the diaphragm deflects. This leads to a change in separation between the plates, causing increase in capacitance. Since the capacitance depends upon separation of overlapping area between plates, the change in capacitance varies with the applied pressure on the diaphragm [5, 6]. The sensitivity of sensor is always influenced by the thickness, shape, and material of diaphragm as well as the type of diaphragm like slotted, bossed or corrugated. The capacitive pressure sensors are widely accepted in design, manufacturing, consumer electronics, medical and automation industries etc. for various applications like tire-pressure monitoring, intraocular pressure measurement, fingerprint application, blood pressure sensing, and barometric pressure measurement etc. [4 - 8].

The blood pressure is measured in form of systolic and diastolic pressure. For blood pressure measurement, systolic and diastolic pressure must be known. The normal blood pressure is 120/80. Where, 120 is systolic and 80 is diastolic pressure. The systolic blood pressure varies normally from 120 mm Hg to 160 mm Hg and diastolic blood pressure varies normally from 80 mm Hg to 100 mm Hg [9].

## 2 Mathematical Modelling

### 2.1 Mathematical modeling of deflection in circular shaped clamped diaphragm

The diaphragm deflection follows Hook's Law i.e. the diaphragm deflection varies linearly when pressure is applied on the diaphragm [10].

If a uniform pressure (P) is applied in normal direction of thin circular flat plate clamped at edges, which is made of elastic, isotropic and homogeneous material, the plate equation can be represented by following third order partial differential equation [11]:

$$\frac{1}{r}\frac{\partial}{\partial r}\left[\frac{1}{r}\frac{\partial}{\partial r}\left(r\frac{\partial w}{\partial r}\right)\right] = \frac{P}{2D} \tag{1}$$

Or,

$$\frac{1}{r}\left[\frac{\partial^3 w}{\partial r^3} + \frac{1}{r}\frac{\partial^2 w}{\partial r^2} - \frac{1}{r^2}\frac{\partial w}{\partial r}\right] = \frac{P}{2D} \tag{2}$$

Where, D = E$t^3$ / $(12 - \text{v}^2)$ is flexural rigidity of the diaphragm, deflection at distance r from the center of the diaphragm is w, thickness of diaphragm is t, Poisson's Ratio is v and Young's modulus of diaphragm material is E. To solve plate equation, the following three boundary conditions are used:

$$w(r = a) = 0 \tag{3}$$

$$\frac{\partial w(r=0)}{\partial r} = 0 \tag{4}$$

$$\frac{\partial w(r=a)}{\partial r} = 0 \tag{5}$$

Here, $a$ is the radius of the diaphragm. Using these three boundary conditions [Eq. (3), (4) and (5)] the deflection in plate can be given by:

$$w(r) = w_{max}\left[1 - \left(\frac{r}{a}\right)^2\right]^2 \tag{6}$$

Where,

$$w_{max} = \frac{Pa^4}{64D} \tag{7}$$

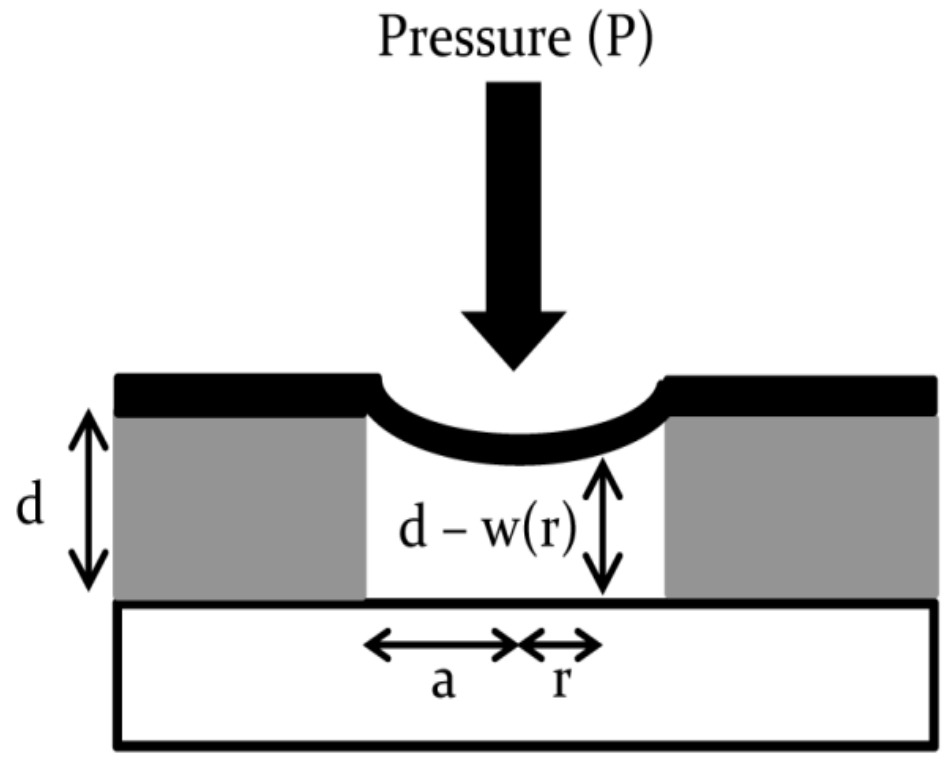

**Fig. 1.** Basic structure of normal mode MEMS capacitive pressure sensor.

### 2.2 Mathematical modeling of normal mode circular shaped capacitive pressure sensor

In this paper, we have proposed an equation of capacitance in a circular diaphragm pressure sensor, when pressure is applied. Based on this equation, the MATLAB® is used for predicting the output of a capacitive sensor. If overlapping area between plates, separation gap between plates and permittivity of medium are A, d and ε, respectively. Then base capacitance can be given by [3, 6]:

$$C_b = \frac{\varepsilon A}{d} \tag{8}$$

And overlapping area for circular plate can be given by:

$$A = \pi a^2 \tag{9}$$

If pressure P is applied on diaphragm, the changed capacitance can be given by [2, 5]:

$$C_w = \int_0^{2\pi} \int_0^a \frac{\varepsilon\, r dr\, d\theta}{d - w(r)} \tag{10}$$

Substituting $w(r)$ from Eq. (6) in Eq. (10), we get -

$$C_w = \frac{2\pi\varepsilon}{d} \int_0^a \frac{rdr}{1 - w_{max}\left[1 - \left(\frac{r}{a}\right)^2\right]^2} \tag{11}$$

After performing integration, we get,

$$C_w = \frac{\varepsilon A}{\sqrt{d\, w_{max}}} \ln\left|\frac{\sqrt{d} + \sqrt{w_{max}}}{\sqrt{d} - \sqrt{w_{max}}}\right| \tag{12}$$

### 2.3 Sensitivity of clamped circular shaped capacitive pressure sensor

We can define the sensitivity of the sensor as the change in capacitance of the sensor in the given pressure range divided by the change in pressure. Sensitivity of sensor can be given by:

$$S = \frac{C_{max} - C_{min}}{P_{max} - P_{min}} \tag{11}$$

## 3 Simulation Results and Discussion

### 3.1 Deflection in diaphragms

Three different diaphragm thicknesses (4 μm, 5 μm and 6 μm) and two different separation gaps (1 μm and 2 μm) are chosen to obtain six different designs. The sensors are designed for a pressure range from 0 to 280 mm Hg above atmospheric pressure (760 mm Hg), in order to measure the blood pressure. This pressure range equals 1.0 to 1.4 bar. In each design, the maximum deflection of diaphragm at maximum pressure (1.4 bar) is kept less than 1/4th of the separation gap to ensure good linearity. Based on this consideration, the diaphragm radius is determined. The deflection in 4 µm, 5 µm and 6 µm thick diaphragms are presented in Figs. 2, 3 and 4, respectively, for two different diaphragm radiuses. These radiuses are obtained for separation gap of 1 µm and 2 µm for each diaphragm thickness. Diaphragm of particular thickness and smaller radius has less than 0.25 µm deflection at 1.4 bar pressure. On the other hand, for the same thickness of diaphragm, the larger radius has deflection less than 0.5 µm at 1.4 bar pressure. The diaphragm deflection increases as radius of diaphragm increases. In all the simulations, silicon is used as the diaphragm material and the following properties are used: E = 169.8 GPa and ν = 0.066.

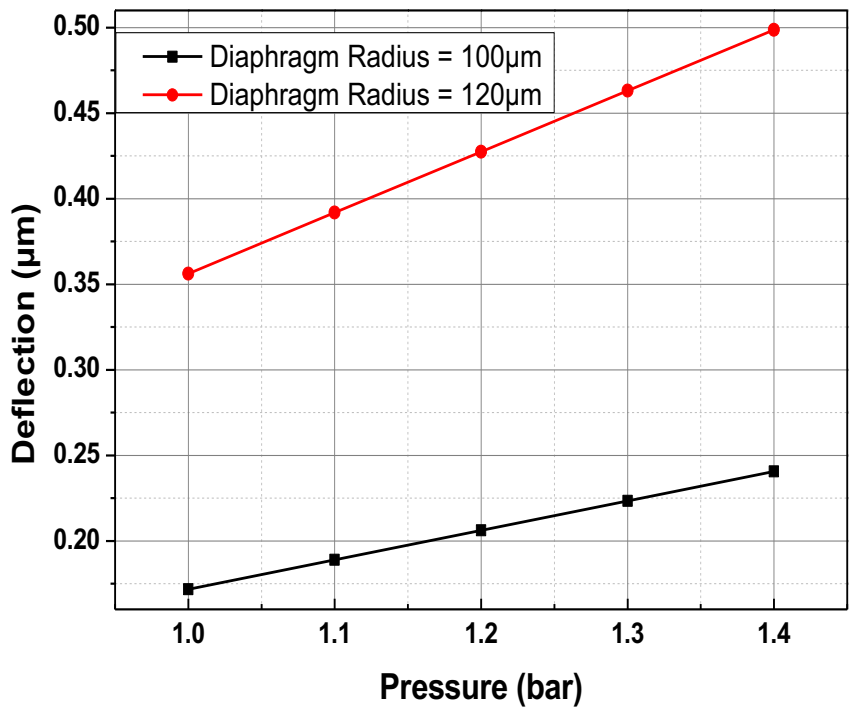

**Fig. 2.** Deflection in 4 µm thick diaphragm .

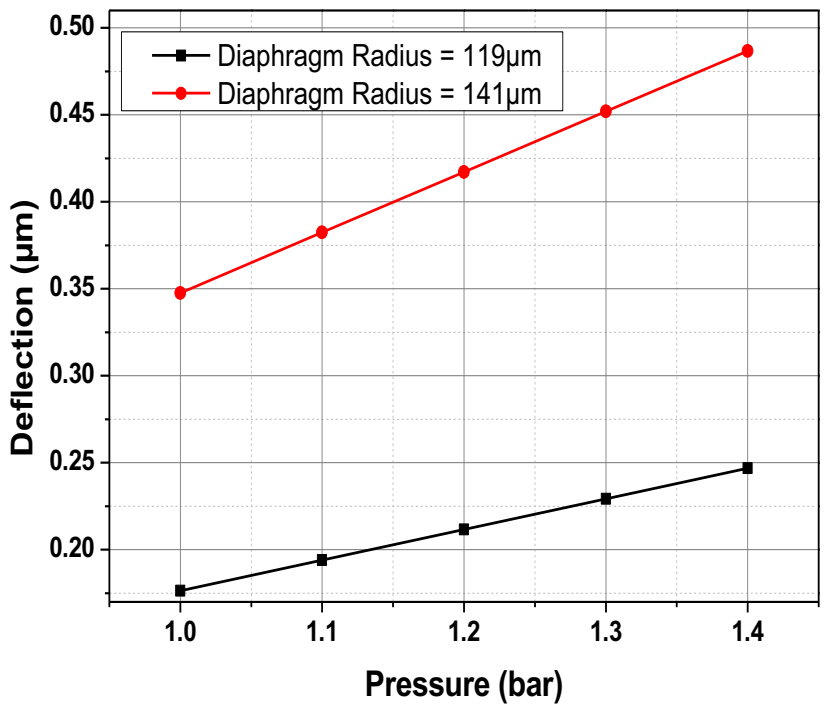

**Fig. 3.** Deflection in 5 µm thick diaphragm.

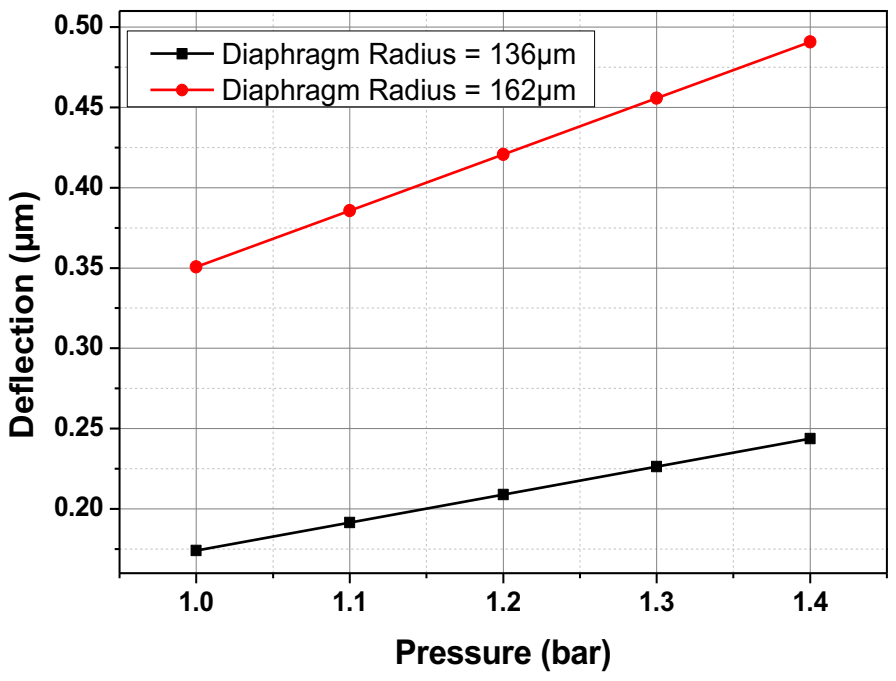

**Fig. 4.** Deflection in 6 µm thick diaphragm.

### 3.2 Base capacitance and capacitance variation with pressure

The base capacitance (or capacitance at 0 bar pressure) for different diaphragm thicknesses and separation are presented in Table I. Base capacitance does not depend on diaphragm thickness but it depends on separation gap and radius of diaphragm.

**Table I.** Base Capacitance for different diaphragm thickness and separation gap.

| Diaphragm Thickness (μm) | Separation Gap (μm) | Diaphragm Radius (μm) | Base Capacitance (pF) |
|---|---|---|---|
| 4 | 1 | 100 | 0.2782 |
| 5 | 1 | 119 | 0.3939 |
| 6 | 1 | 136 | 0.5145 |
| 4 | 2 | 120 | 0.2003 |
| 5 | 2 | 141 | 0.2765 |
| 6 | 2 | 162 | 0.365 |

In the present work, pressure range varies between 1.0 to 1.4 bar. The graphs between pressure and capacitance for the six different designs are shown in Figs. 5 to 7.

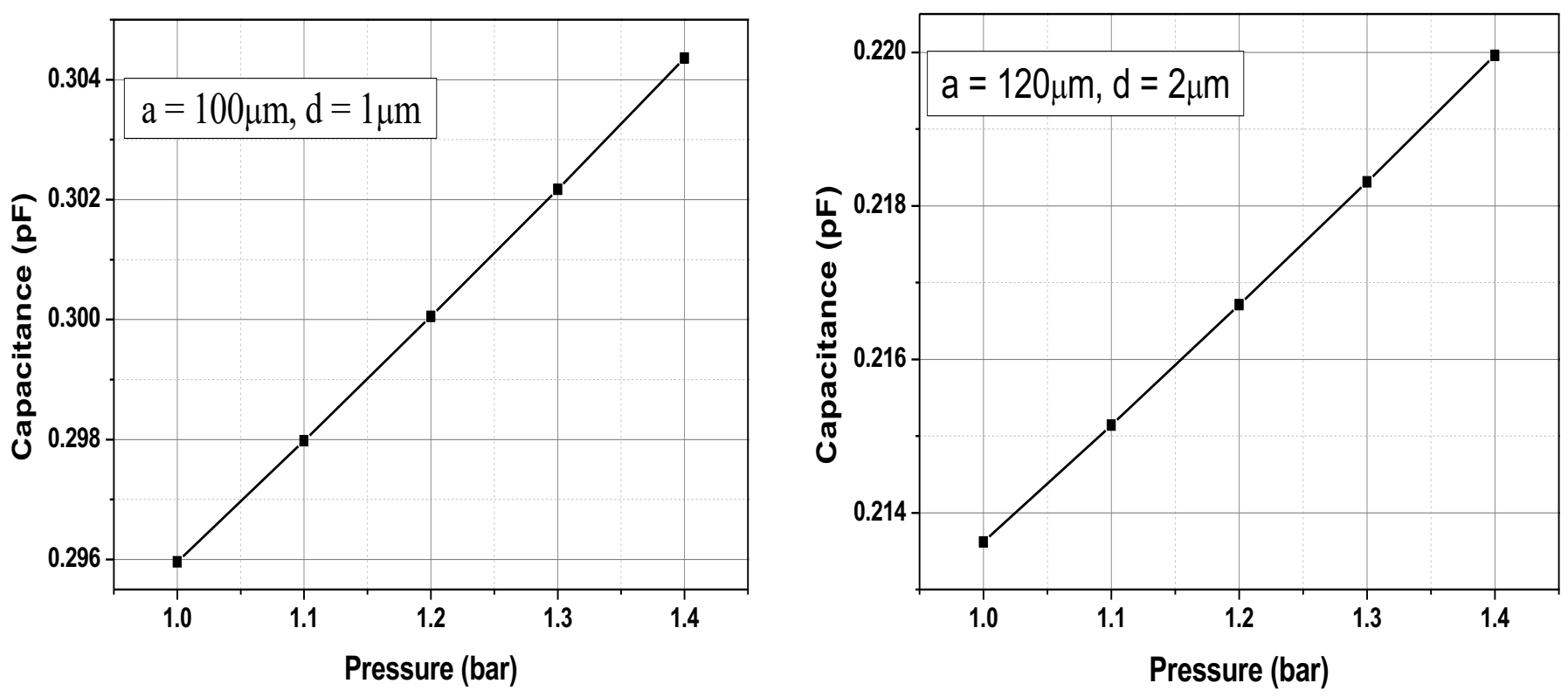

**Fig.5.** Capacitance vs. Pressure for 4 μm thick diaphragm.

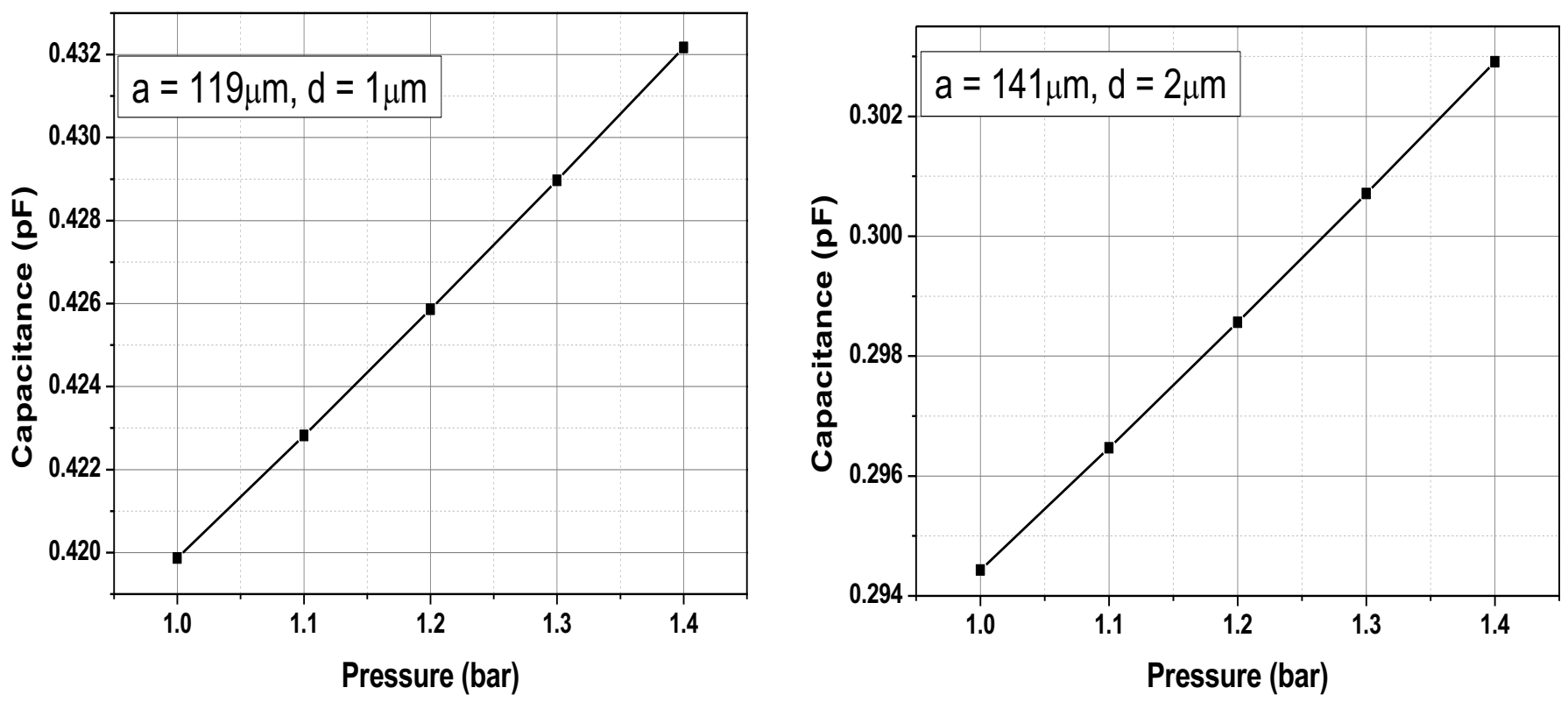

**Fig. 6.** Capacitance vs. Pressure for 5 μm thick diaphragm.

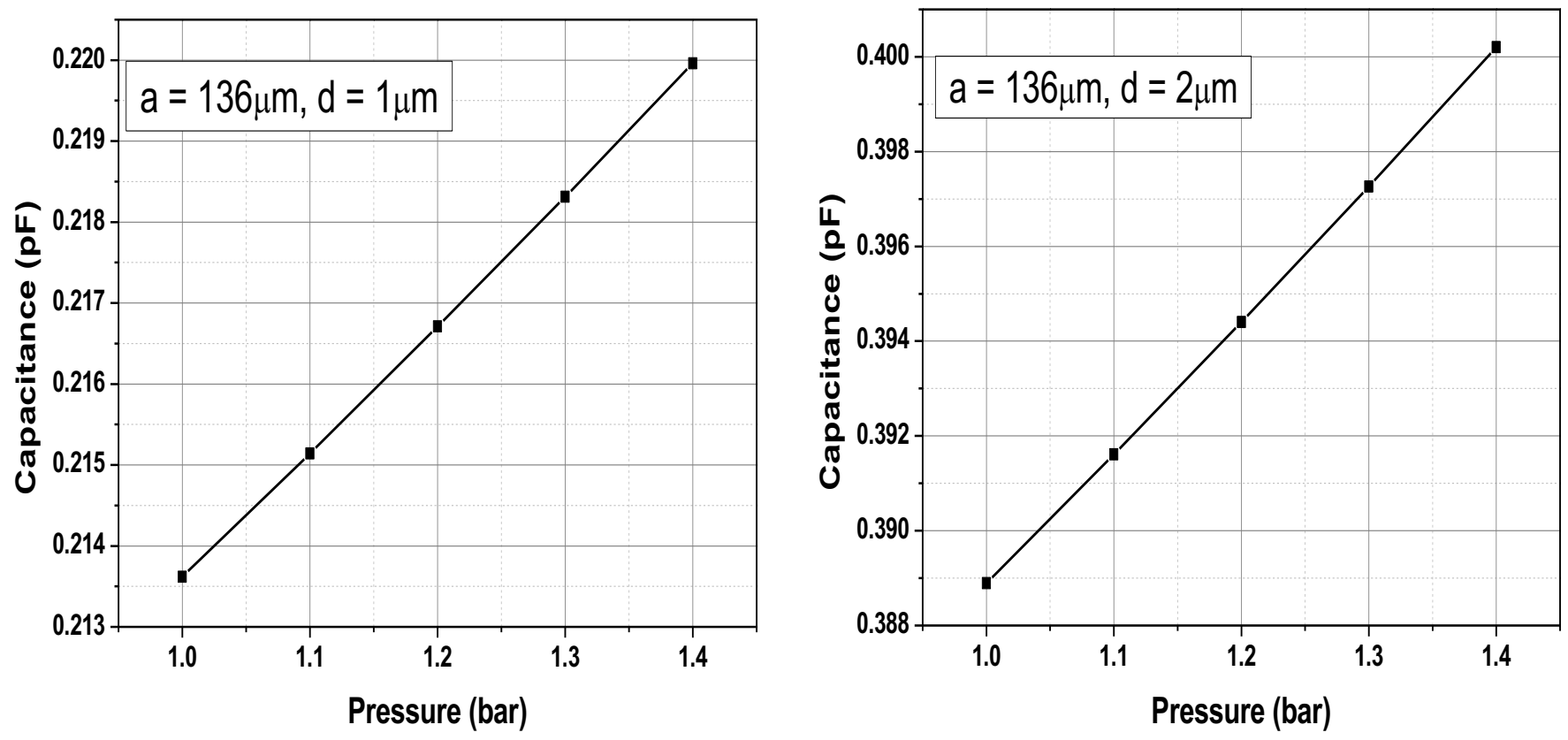

**Fig. 7.** Capacitance vs. Pressure for 6 μm thick diaphragm.

### 3.3 Sensitivity

The sensitivity of the six different designs is shown in Table II.

**Table II.** Sensitivity of different designs.

| Diaphragm Thickness (μm) | Diaphragm Radius (μm) | Separation gap (μm) | Sensitivity (fF/bar) |
|---|---|---|---|
| 4 | 100 | 1 | 21.0 |
| 4 | 120 | 2 | 15.85 |
| 5 | 119 | 1 | 30.75 |
| 5 | 141 | 2 | 21.2 |
| 6 | 136 | 1 | 39.52 |
| 6 | 162 | 2 | 28.28 |

## 4 Conclusions

In this paper, capacitive pressure sensor is utilized for blood pressure monitoring. In the designs, deflection in diaphragm is less than one fourth of separation between plates. Using mathematical modeling, the solution of capacitance after applying pressure is obtained which is the logarithmic function. The advantage of capacitive sensor is high pressure sensitivity and insensitivity to temperature change. However, loss in signal due to parasitic capacitance and low change in capacitance are major disadvantages. The circuitry used for measurement must address these issues.

From Table II, we can conclude that the design having a diaphragm thickness of 6 μm, separation gap of 1 μm and diaphragm radius of 136 μm has the maximum sensitivity amongst all the six designs. It has a sensitivity of 39.52 fF/bar in the required pressure range. Therefore, this design is best suited for the measurement of blood pressure. The designed sensor can be fabricated with appropriate fabrication process sequence.

## Acknowledgement

Authors would like to acknowledge the generous support of Director, CSIR – Central Electronics Engineering Research Institute (CEERI), Pilani. They would like to thanks all the scientific and technical staff of Process Technologies Group-SSA for their support and co-operation.